\documentclass[aps,pra,twocolumn,showpacs,preprintnumbers,prb,superscriptaddress]{revtex4-1}
\usepackage{graphicx}  
\usepackage{dcolumn}   
\usepackage{bm}        
\usepackage{amssymb}   
\usepackage{framed}
\usepackage{amsmath}
\usepackage{hhline}
\usepackage{color}
\definecolor{bluegreen}{rgb}{0,0.2,0.8}
\usepackage{subfigure,amsmath,verbatim,moreverb}
\usepackage{tabularx}
\usepackage{adjustbox}
\usepackage{lipsum}
\usepackage{longtable}
\usepackage{booktabs}
\usepackage{latexsym}
\usepackage{dcolumn}
\usepackage{amsmath}
\usepackage{epsf}
\usepackage{float}
\usepackage{hyperref}

\usepackage{longtable}
\usepackage{booktabs}
\usepackage{threeparttable}

\newcommand{\R}{\mathbf{r}}

\usepackage{etoolbox}
\AtBeginEnvironment{align}{\setcounter{subeqn}{0}}
\newcounter{subeqn} %

\begin{document}
\title{Simple and accurate screening parameters for dielectric-dependent hybrids}
\author{Subrata Jana}
\email{subrata.niser@gmail.com}
\affiliation{Department of Chemistry \& Biochemistry, The Ohio State University, Columbus, OH 43210, USA}
\author{Arghya Ghosh}
\affiliation{Department of Physics, Indian Institute of Technology, Hyderabad, India}
\author{Lucian A. Constantin}
\affiliation{Istituto di Nanoscienze, Consiglio Nazionale delle Ricerche CNR-NANO, 41125 Modena,Italy}
\author{Prasanjit Samal}
\affiliation{School of Physical Sciences, National Institute of Science Education and Research, HBNI,
Bhubaneswar 752050, India}

\date{\today}

\begin{abstract}

A simple effective screening parameter for screened range-separated hybrid is constructed from the compressibility sum rule in the context of linear-response time-dependent Density Functional Theory. When applied to the dielectric-dependent hybrid (DDH), it becomes remarkably accurate for bulk solids compared to those obtained from fitting with the model dielectric function or depending on the valence electron density of materials. The present construction of the screening parameter is simple and realistic. The screening parameter developed in this way is physically appealing and practically useful as it is straightforward to obtain using the average over the unit cell volume of the bulk solid, bypassing high-level calculations of the dielectric function depending on random-phase approximation. Furthermore, we have obtained a very good accuracy for  energy band gaps, positions of the occupied $d-$ bands, ionization potentials, optical properties of semiconductors and insulators, and geometries of bulk solids (equilibrium lattice constants and bulk moduli) from the constructed DDH.

\end{abstract}

\maketitle

\section{Introduction}

Kohn-Sham (KS) density functional theory (DFT)~\cite{kohn1965self,hohenberg1964inhomogeneous} becomes the state-of-the-art
method for the electronic structure calculations of solids and materials~\cite{burke2012perspective,engel2013density,Jones2015,CoheMoriYang2012,HasnRefsProb2011,kummel2008orbital,dftsharing2022}. Although it
is an exact theory, one must approximate the exchange-correlation (XC) part of
KS potential, which includes all the many-body interactions beyond the Hartree
theory. The development of new XC approximations having insightful physical
content and which are also accurate as well as efficient for solids is
always desirable~\cite{perdew2001jacob,scuseriaREVIEW05,della2016kinetic,perdew2008density,perdew2005prescription}. In this respect, semilocal XC approximations~\cite{perdewPRL96,perdew2008restoring,lee1988development,PerdewWorkhorse2009,taoPRL03,tao2016accurate,patra2019relevance,constantin2016semilocal,constantin2016hartree,constantin2015gradient,constantin2011correlation,sun2015strongly,furness2020accurate,mejia2020metagga,jana2021szs,jana2019improving,patra2020way,jana2021accurate} are quite useful,
because of their efficiency
~\cite{singh2009surface,patra2017properties,jana2018assessing,HaasTranBlaha2009,sun2011self,tran2016rungs,mo2017assessment,peng2017synergy,zhang2017comparative,shani2018accurate,JanaSharmaSamal2018,patra2021correct,jana2020insights,ghosh2021improved}. However, there are limitations when applied these semilocal XC functionals to calculate energy gaps for solids~\cite{perdew2017understanding,tran2007band,Borlido2020,patra2019efficient,fabien2018assessment,
 jana2018assessing,tran2017importance,fabien2019semilocal}, optical spectrum~\cite{paier2008dielectric,wing2019comparing,stadele1999exact,
petersilka1996excitation,kim2002excitonic,terentjev2018gradient,sharma2011bootstrap,
rigamonti2015estimating,van2002ultranonlocality,
cavo2020accurate,jana2020improved,OhadWingGant2022,WingHabeJonah2019,AshwinDahvydLeeor2019}, and semiconductor
defects~\cite{DeakLorkeAradi2019,LewiMatsBell2017,BatiEnriHeyd2006,DeakAradThom2010,RauchFranmigu2021}. All these deficiencies of the semilocal XC functionals are related to the
known de-localization error~\cite{perdew2017understanding,ShuTru2020}, which leads to the construction of hybrid functions
with fractions of HF mixing~\cite{heyd2003hybrid,krukau2006influence,tao2008exact,heyd2004efficient,jana2020screened,jana2020improved,jana2022solid,janameta2018,patralong2018,jana2019long,jana2019screened,jana2018many,Garrick2020exact}. Although the hybrid XC approximations of DFT solve many
problems, they also have some limitations when a fixed HF mixing is used~\cite{DeakLorkeAradi2019,WingHabeJonah2019,Wang2016hybrid}
Nowadays, hybrid functionals with system-dependent HF mixing are fairly popular methods. Those are known as the dielectric-dependent hybrids (DDHs)~\cite{ShimNaka2014,SkonGovoGall2014,BrawVorosGovo2016,SkonGovoGall2016,WeiGiaRigPas2018,CuiWangZhang2018,MichPeteThom2020}, where the HF
mixing is proportional to the inverse of the macroscopic static dielectric
constant of the system under study. Such hybrids are developed and applied to
solids for quite some time~\cite{GeroBottCaraOnid2015,GeroBottCara2015,MicelChenIgor2018,ZhengGovoniGalli2019,GeroBottValeOnid2017,HinuKumaTana2017,BrawGovoVoro2017,LiuCesaMart2020}. DDHs can be considered as the higher rung hybrids
than those proposed from regular fixed HF mixing. Also, DDHs are
computationally more expensive than regular fixed HF mixed hybrids, in
the sense that one needs to calculate the dielectric constant of the material
beforehand. However, the great advantages of DDHs are that they are constructed
smartly using the same philosophy as COH-SEX (local Coulomb hole plus screened
exchange)~\cite{AndrKressHinu2014} by fulfilling many important constraints that the exact XC functional must observe. Therefore, those possess similar accuracy as $GW$ for band gaps
and Bethe-Salpeter equation (BSE) for optical spectra~\cite{wing2019comparing}.
Regarding the several recently proposed DDHs, we recall range-separated-DDH (RS-
DDH)~\cite{BrawVorosGovo2016,SkonGovoGall2016}, DDH based on the Coulomb-attenuated method (DD-RSH-CAM)~\cite{WeiGiaRigPas2018},
and doubly-screened DD hybrid (DSH)~\cite{CuiWangZhang2018} based on their range separation.
Also, there are other ways of implementing the DDHs such as satisfaction of the
Koopmans-theorem~\cite{MichPeteThom2020}. A fairly good description and comparison of different
versions of hybrids are discussed in ref.~\cite{LiuCesaMart2020}. Although the system-dependent macroscopic dielectric constant for DDHs is calculated from first principles (such as Perdew-Burke-Ernzerhof (PBE) or random-phase approximation (RPA) on the top of the PBE calculations (RPA@PBE)), the
screening parameters are constructed from several philosophies. Such as from the fitting of the long-wavelength limit of highly accurate dielectric
functions~\cite{WeiGiaRigPas2018,wing2019comparing,LiuCesaMart2020} calculated from random-phase approximation (RPA) or nanoquanta
kernel and partially self-consistent $GW$ calculations~\cite{wing2019comparing,LiuCesaMart2020} or from valance electron
density~\cite{BrawVorosGovo2016,SkonGovoGall2016,CuiWangZhang2018,MichPeteThom2020}. Both are non-empirical choices and required no optimization
procedure.
In this work, we propose an alternative procedure for obtaining the range-separated parameter for DDHs using a simple and effective way via the
compressibility sum rule, which connects the screening parameter with the
exchange energy density. It is quite a realistic way of obtaining the
screened parameter, where the relationship can be established through the
linear-response time-dependent DFT (TDDFT). Importantly, the present
construction gives a very realistic result similar to those obtained from the model
dielectric function. We
assess the accuracy of screening parameters with DD-RSH-CAM~\cite{WeiGiaRigPas2018} for the electronic
properties of solids, especially energy gaps, geometries, and optical
properties.
The rest of the paper is organized as follows. Section~\ref{secii} describes the
generalized formulation of the range-separated DDH along with the construction
of the screening parameter developed in this work. Section~\ref{seciii} presents
results obtained from non-empirical screening parameters using the DD-RSH-CAM
for solid properties. Section~\ref{seciv} summarizes and concludes the work of
this paper.

\section{Theory}
\label{secii}

\subsection{Generalized form}

We start from the Coulomb attenuated method (CAM) style ansatz of the screened-range-separated hybrid (SRSH) functional by partitioning the Coulomb interaction as~\cite{kronik2018dielectric},
\begin{equation}
\frac{1}{r}=\frac{\alpha+\beta~erf(\mu r)}{r}+\frac{1-[\alpha+\beta~erf(\mu r)]}{r}~,
\label{eqsec1-1}
\end{equation}
where $\mu$ is the range-separation parameter. In SRSH, the first term is treated by a Fock-like operator. The second term is treated by semilocal exchange, which is based on the semilocal (SL) GGA functional (PBE) in the present case. However, meta-GGA semilocal functionals can also be used~\cite{jana2022solid}. Following Eq.~\ref{eqsec1-1} the expression of the exchange-correlation (XC) functional becomes,
\begin{equation}
\begin{split}
 E_{xc}^{SRSH}=(1-\alpha)E_x^{SR-SL,\mu}+\alpha E_x^{SR-HF,\mu}\\
            +[1-(\alpha+\beta)]E_x^{LR-SL,\mu}+(\alpha+\beta)E_x^{LR-HF,\mu}+E_c^{SL}~.
 \label{eqsec1-2}
\end{split}
 \end{equation}
Here, $\alpha$ and $\beta$ control the fraction of short-range and long-range exchange to the above decomposition. $\mu$ is the range-separation parameter. The aforementioned generalized decomposition can take several forms depending on the tuning of $\alpha$ and $\beta$ parameters. For example, (i) with $\beta=-\alpha$, the screened hybrid with SR-HF and LR-SL is recovered. This type of hybrid is useful for solids~\cite{heyd2003hybrid,jana2020screened,jana2020improved}, (ii) with the choice of $\alpha+\beta=1$, in LR, the HF is always recovered~\cite{VydrScuse2006,jana2019long}. This type of hybrid is useful for finite systems, especially for the long-range excitation of molecules~\cite{Kronik2012excitation}, and finally the (iii)
the global hybrid functional is obtained by considering $\beta=0$~\cite{PBE0}.

Though choice (i) is quite convenient for solids and popularly used in the name of Heyd-Scuseria-Ernzerhof (HSE)~\cite{heyd2003hybrid}, it underestimates the band-gaps of insulators~\cite{tran2017importance} and defect formation energies~\cite{DeakLorkeAradi2019} because of the lack of dielectric screening. Considering this limitation, the SRSH hybrid has been constructed by proposing the dielectric screening of solids as $\alpha+\beta=\epsilon^{-1}_{\infty}$. This SRSH functional has the following expression
\begin{equation}
\begin{split}
 E_{xc}^{SRSH}=(1-\alpha)E_x^{SR-SL,\mu}+\alpha E_x^{SR-HF,\mu}\\
            +(1-\epsilon^{-1}_{\infty})E_x^{LR-SL,\mu}+\epsilon^{-1}_{\infty}E_x^{LR-HF,\mu}+E_c^{SL}~,
 \label{eqsec1-3}
\end{split}
 \end{equation}
where $\epsilon^{-1}_{\infty}$ is the inverse of the macroscopic static dielectric constant which is material specific. The main motivation of the underlying  approximation is followed from Green's function based many-body approaches ($GW$ exchange-correlation self-energy methods, $\Sigma_{xc}$), where local Coulomb hole (COH) plus screened exchange (SEX) (COHSEX) are taken into account. (See ref.~\cite{CuiWangZhang2018} for the connection between COHSEX and DDHs.) The corresponding potential of the screened exchange is given as,
 \begin{eqnarray}
 V_{xc}^{SRSH}({\bf {r,r'}})&=&[\alpha-(\alpha-\epsilon^{-1}_{\infty})erf(\mu r)]V_x^{HF}({\bf {r,r'}})\nonumber\\
 &-&(\alpha-\epsilon^{-1}_{\infty})V_x^{SR-SL,\mu}({\bf {r}})+(1-\epsilon^{-1}_{\infty})V_x^{SL}({\bf {r}})\nonumber\\
 &+& V_c({\bf {r}})~,
 \label{eqsec1-4}
 \end{eqnarray}
where $V_x^{HF}({\bf {r,r'}})$ is the full-range HF exchange,
\begin{equation}
 V_x^{HF}({\bf {r,r'}}) = -\sum_{n{\bf{k}}}w_{n{\bf{k}}}f_{n{\bf{k}}}\frac{\phi^{KS}_{n{\bf{k}}({\bf{r'}})}\phi^{KS}_{n{\bf{k}}}({\bf{r}})}{|{\bf{r}}-{\bf{r}'}|}~,
 \label{eqsec1-5}
\end{equation}
 $\phi^{KS}_{n{\bf{k}}}$'s are the KS orbitals or basis, $V_x^{SL}({\bf {r}})$ is the full-range semilocal functional, which is PBE functional in the present case, $V_x^{SR-SL,\mu}({\bf {r}})$ is the short-range (SR) part of the PBE functional, and $V_c({\bf {r}})$ is the PBE correlation.
It may be noted that for solids the reciprocal space representation of $erf(\mu r)$ becomes $e^{-|\bf{G}|^2/(4\mu)}$, where $\bf{G}$ is the reciprocal lattice vector. Several choices of the $\alpha$, $\epsilon^{-1}$, and $\mu$ exist and based on those choices rungs of screened exchange or DDHs functionals may be constructed (see ref.~\cite{LiuCesaMart2020} for details).

In particular, we consider the case $\alpha=1$ which is used in the doubly screened hybrid (DSH) of Ref.~\cite{CuiWangZhang2018} and in the dielectric-dependent range-separated hybrid functional based on the Coulomb-attenuating method (CAM) (DD-RSH-CAM)~\cite{WeiGiaRigPas2018}. In the later case, the model dielectric function is
\begin{equation}
 \varepsilon^{-1}({\bf{G}})=1-(1-\epsilon^{-1}_{\infty})e^{-|\bf{G}|{^2}/(4\mu)}.
 \label{eqsec1-6}
\end{equation}
However, the accuracy of Eq.~\ref{eqsec1-3} depends on two main aspects:

(i) The macroscopic static dielectric constant, $\epsilon^M_\infty$:  is mostly calculated in a first-principles way using the linear response TDDFT method~\cite{Souza2002first,Nunes2001berry,Gajdo2006linear}. The RPA@PBE dielectric constants are reasonably well described~\cite{WeiGiaRigPas2018}. For hybrids or DDHs, the dielectric constants are calculated using RPA+$f_{xc}$, where $f_{xc}$ is the XC kernel~\cite{WeiGiaRigPas2018,LiuCesaMart2020}. Although there are several XC kernels (see ref.~\cite{Olsen2019beyond} for a review), we recall that for DD-RSH-CAM, the bootstrap approximation of Sharma et. al.~\cite{sharma2011bootstrap} ($f_{xc}^{bootstrap}$) is used.

(ii) The screening parameter, $\mu$: is obtained following several procedures such as: (a) from the fitting of the model dielectric function (Eq.~\ref{eqsec1-6}) with the long-wavelength limit of the diagonal elements of  dielectric function i.e., $\varepsilon^{-1}_{{\bf{G,G}}}({\bf{q}}\to 0, \omega=0)$ obtained from RPA calculations as mentioned in ref.~\cite{WeiGiaRigPas2018}. However, to obtain an accurate dielectric function, one needs additional calculations of highly accurate RPA (and/or ``nanoquanta'' kernel combined with partially self-consistent $GW$)~\cite{WeiGiaRigPas2018}. (b) Alternatively, the screening parameter, $\mu$ can also be obtained from the valence
electron density of the system as referred in Refs.~\cite{BrawVorosGovo2016,SkonGovoGall2016,CuiWangZhang2018}.

In the following, we propose a simple expression for the screening parameter $\mu$, derived from the linear response TDDFT approach.

\begin{figure}
\begin{center}
\includegraphics[width=\columnwidth]{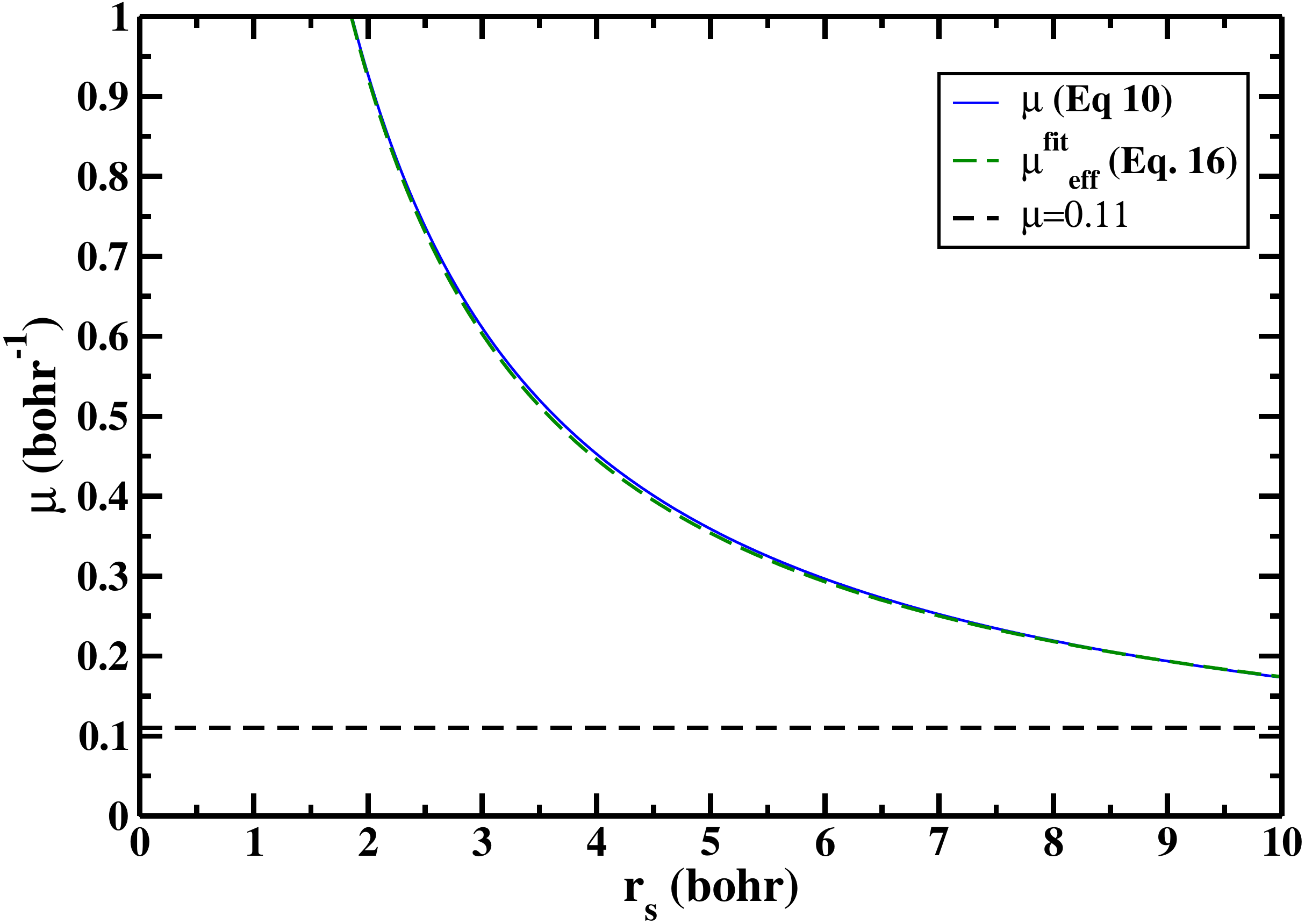}
\end{center}
\caption{\label{fig1} Shown is the $\mu(r_s)$ as a function of local Seitz radius, $r_s$ for $0<r_s<10$ bohr.
}
\end{figure}
\begin{figure}
\begin{center}
\includegraphics[width=\columnwidth]{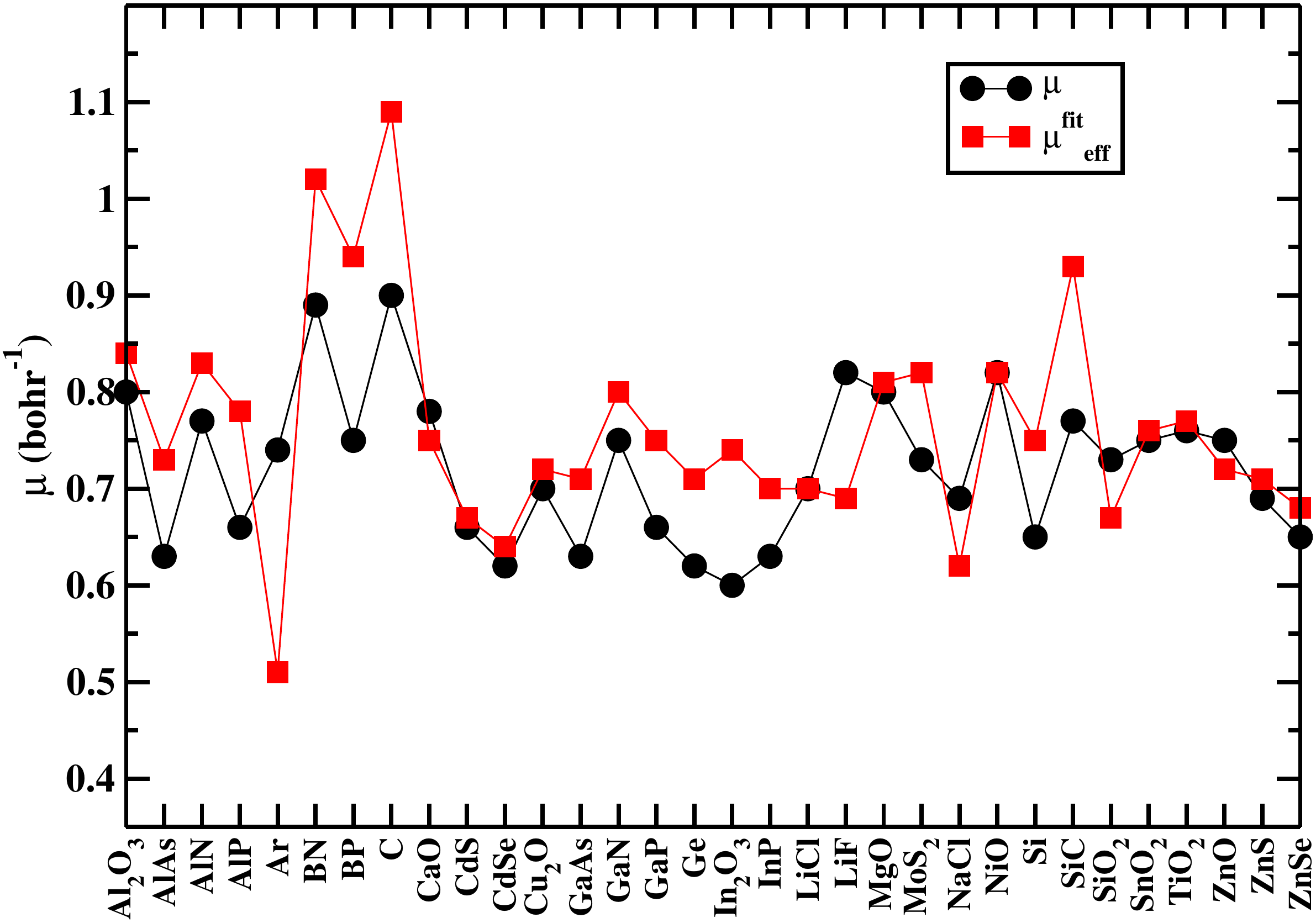}
\end{center}
\caption{\label{fig2} Comparison of $\mu$ values for several solids.}
\end{figure}
\begin{figure}
\begin{center}
\includegraphics[width=\columnwidth]{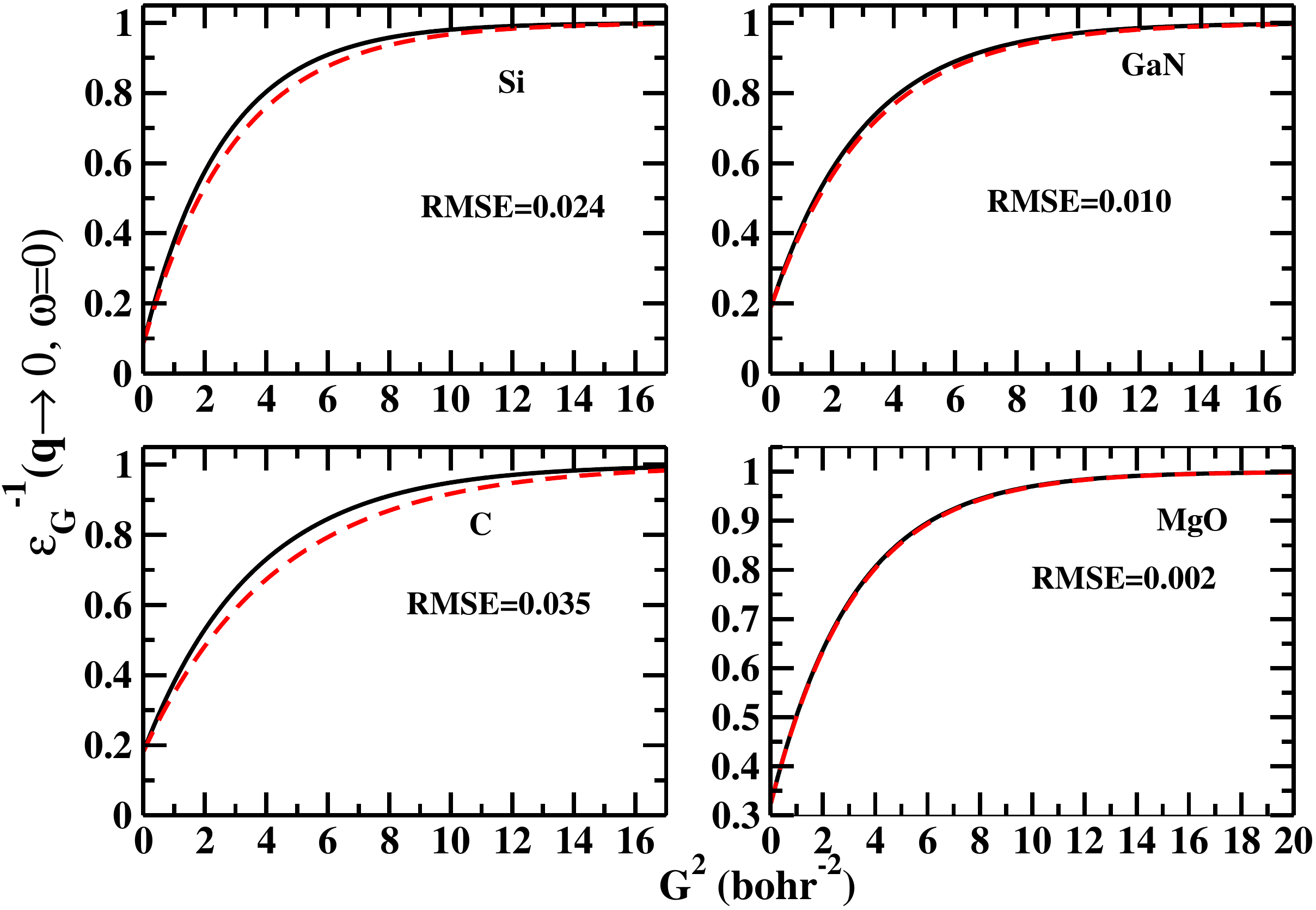}
\end{center}
\caption{\label{fig3} Comparison of model dielectric function as a function of ${\bf{G}}$ for four solids. Black solid line for the $\mu$ values of ref.~\cite{WeiGiaRigPas2018}, whereas red dotted line for $\mu_{eff}^{fit}$. Relative mean square error (RMSE) of the model dielectric function as obtained from $\mu_{eff}^{fit}$ from the actual one (obtained using $\mu$) is also shown.}
\end{figure}

\subsection{Construction of the screening parameter}

In the linear-response TDDFT, the interacting ($\chi(\R,\R';\omega)$) and
noninteracting ($\chi_0(\R,\R';\omega)$) density-density response functions are connected by the following Dyson like equation~\cite{Gross1996},

\begin{eqnarray}
\chi({\bf{r}},{\bf{r'}};\omega)=\chi_0({\bf{r}},{\bf{r'}};\omega)&+&\int d{\bf{r}}_1d{\bf{r}}_2~\chi_0({\bf{r}},{\bf{r'}};\omega)\nonumber\\
&\times& v_{eff}[n]({\bf{r}},{\bf{r'}};\omega)\chi({\bf{r}},{\bf{r'}};\omega)~,
\label{sc-eq1}
\end{eqnarray}
where
\begin{equation}
v_{eff}[n]({\bf{r}},{\bf{r'}};\omega)=\frac{1}{|{\bf{r}}-{\bf{r}}'|}+f_{xc}[n]({\bf{r}},{\bf{r'}};\omega)
\label{sc-eq2}
\end{equation}
is Coulomb plus XC kernel known as the effective potential. If $f_{xc}[n]({\bf{r}},{\bf{r'}};\omega)$ is zero, then the  RPA~\cite{HarrGriff1975,LangPerd1977} is recovered. Therefore, $f_{xc}[n]({\bf{r}},{\bf{r'}};\omega)$ should account for the short-range correlation, which is missing in RPA. Following these considerations, Constantin and Pitarke (CP) proposed a simple and accurate approximation of
$v_{eff}[n]({\bf{r}},{\bf{r'}};\omega)$ for the three-dimensional (3D) uniform electron gas (UEG) \cite{constantin2007simple,constantin2016simple}
\begin{equation}
 v^{CP}_{eff}[n]({\bf{r}},{\bf{r'}};\omega) = \frac{{\rm erf}(|{\bf{r}}-{\bf{r}}'|/\sqrt{4k_{n,\omega}})}{|{\bf{r}}-{\bf{r}}|}.
 \label{sc-eq3}
\end{equation}

Note that we already consider this type of splitting in the Coulomb interaction of DDH construction. Also, $4k_{n,\omega}$ is a frequency and density-dependent function that controls the long-range effects of bare Coulomb interaction. Therefore, a direct connection between $4k_{n,\omega}$ and screened parameter $\mu$ can be established as follows,
\begin{eqnarray}
 \mu=\mu_{eff} &=& \frac{1}{\sqrt{4k_{n,\omega=0}}}~.
 \label{eq-theo-secb-8}
\end{eqnarray}
Here, we denote $\mu_{eff}$ as the ``{\it{effective}}'' screening to distinguish it from the actual screening parameter used in the DD-RSH-CAM.
Note that for the DDH functionals of the ground-state DFT, we have to consider the static ($\omega=0$) case.

The XC kernel for 3D UEG can be derived using the Fourier transform of Eqs.~\ref{sc-eq2} and ~\ref{sc-eq3} as \cite{constantin2007simple}
\begin{equation}
 f_{xc}(n;q,\omega)=\frac{4\pi}{q^2}[e^{-k_{n,\omega}q^2}-1]~.
 \label{sc-eq4}
\end{equation}
Thus in the long-wavelength ($q\to 0$) limit, one can obtain,
\begin{equation}
k_{n,\omega}=-\frac{1}{4\pi} f_{xc}(n;q\to 0,\omega)~.
  \label{eq-theo-secb-10}
\end{equation}
On the other hand, the long-wavelength limit of the static XC kernel $f_{xc}(n;q\to 0,\omega=0)$ satisfies the compressibility sum rule~\cite{Ichimaru1982strongly},
 \begin{equation}
  f_{xc}(n;q\to 0,\omega=0)=\frac{d^2}{dn^2}(n\epsilon_{xc}(n))
  \label{eq-theo-secb-11}
 \end{equation}
where $\epsilon_{xc}(n)$ is the XC energy per particle of the 3D UEG. We use the Perdew and Wang parametrization of the local density approximation (LDA) correlation energy per particle \cite{PerdWang1992}. Note that the LDA XC kernel is remarkably accurate for $q < 2q_F$ ($q_F=(3\pi^2n)^{1/3}$ being the Fermi wavevector), which explains the success of LDA (and semilocal functionals that recover LDA for the 3D UEG) for bulk solids~\cite{PerdLangSahni1977,MoroCepeSena1995}.

Finally, Eqs.~\ref{eq-theo-secb-10} and ~\ref{eq-theo-secb-11} give
\begin{eqnarray}
 k_{n,\omega=0} = -\frac{1}{4\pi}\frac{d^2}{dn^2}[n\epsilon_{xc}(n)]~,
 \label{eq-theo-secb-12}
 \end{eqnarray}
and $\mu_{eff}$ can be found from Eq.~\ref{eq-theo-secb-8}. Noteworthy, Eq.~\ref{eq-theo-secb-8} is the central equation of this paper, which establishes a direct connection between the screening parameter and the LDA XC energy per particle ( $\epsilon_{xc}(n)$ )\cite{PerdWang1992} which depends only on the local Seitz radius, $r_s=(\frac{3}{4\pi n})^{1/3}$ and the relative spin polarization, $\zeta=\frac{n_{\uparrow}-n_{\downarrow}}{n}$.

To simplify the computational implementation for bulk solids, we consider the average of $r_s$ over the unit cell volume, $\Omega_{cell}$ as,
\begin{equation}
 \langle r_s \rangle=\frac{1}{\Omega_{cell}}\int_{cell} \Big(\frac{3}{4\pi(n_{\uparrow}({\bf{r'}})+n_{\downarrow}({\bf{r'}}))}\Big)^{1/3}~d^3r'~.
 \label{eq-theo-secb-13}
\end{equation}
We recall that this averaging technique over the unit cell has been considered earlier, e.g. in the construction of the modified Becke-Johnson (MBJ) semilocal exchange potential ~\cite{Tran2009accurate,Borlido2020exchange,Rauch2020accurate,Rauch2020local,Tran2021bandgap,patra2021efficient}, local range separated hybrid functionals~\cite{Marques2011density}, and XC kernel for optical properties of semiconductors ~\cite{terentjev2018gradient}.
Thus, for computational simplicity, we fit the exact $\mu_{eff}(r_s)$ curve with the following formula,
\begin{equation}
 \mu_{eff}^{fit}= \frac{a_1}{\langle r_s \rangle} + \frac{a_2 \langle r_s \rangle}{1 + a_3 \langle r_s \rangle^2}~,
\end{equation}
where $a_1= 1.91718$, $a_2= -0.02817$, and $a_3=0.14954$.

In Fig.~\ref{fig1}, we plot $\mu(r_s)$ versus $r_s$ for $0<r_s<10$ bohr. As one can see, $\mu_{eff}(r_s)$ and $\mu_{eff}^{fit}$ agree very well, the curves being almost indistinguishable. We also observe that $\mu_{eff}$
is significantly bigger than the HSE one ($\mu^{HSE}=0.11$), and only in the low-density regime ($r_s>10$) they become comparable. As a side note, $\mu_{eff}(r_s)$ seems also very realistic for LC-type hybrid functionals, where $\mu^{LC} \approx 0.50$ bohr$^{-1}$, (see for example Table IV of Ref. \cite{jana2019long}), because
$\mu_{eff}(r_s) > 0.5$ bohr$^{-1}$ for $r_s < 3.6$ bohr and $\mu_{eff}(r_s) < 0.50$ bohr$^{-1}$ for $r_s > 3.6$ bohr.

\section{Results}
\label{seciii}

\begin{figure}
\begin{center}
\includegraphics[width=\columnwidth]{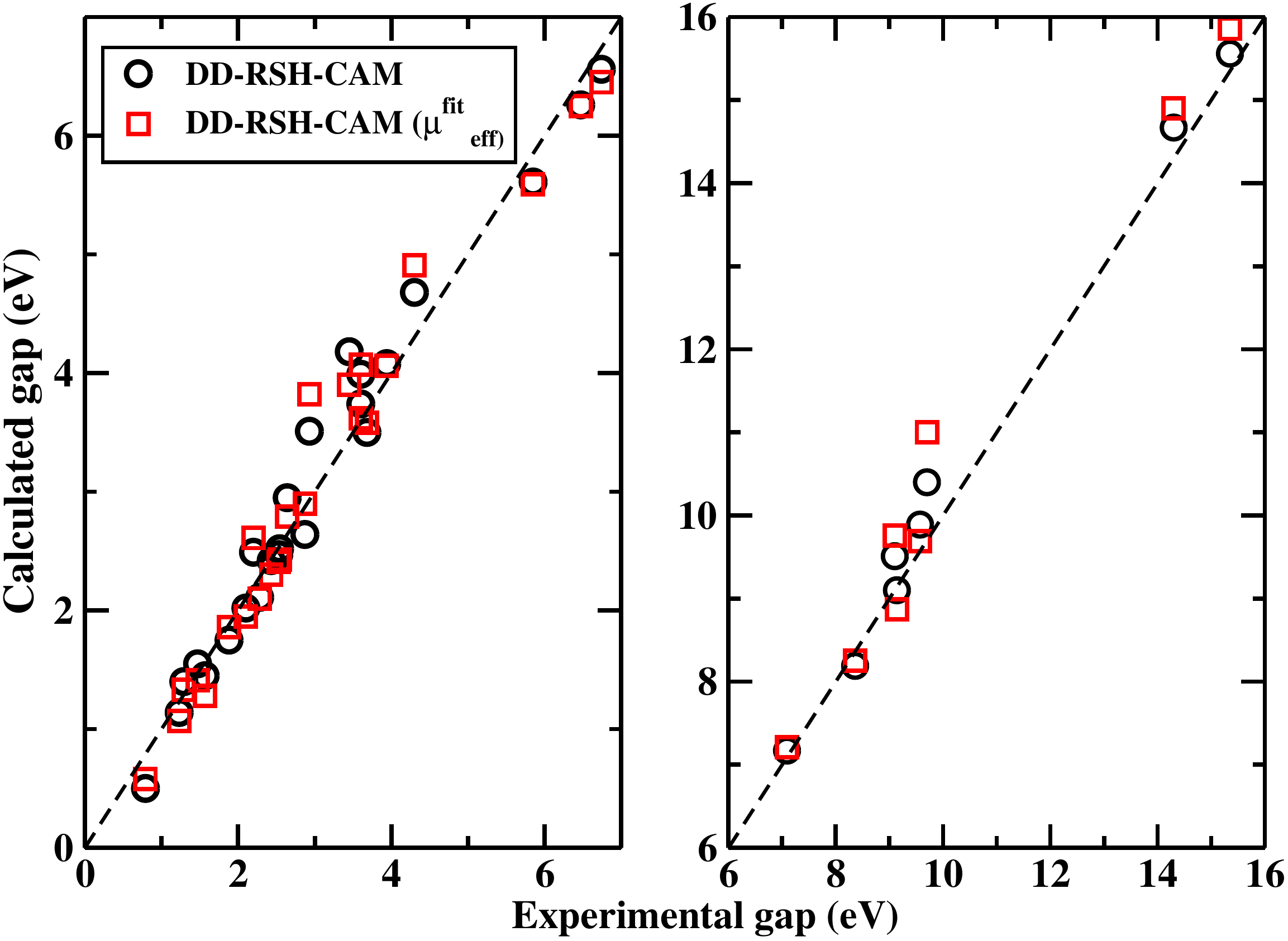}
\end{center}
\caption{\label{fig4} Shown is the calculated versus experimental band gaps of $32$ solids using different methods. $\epsilon_\infty^{-1}$ obtained from DD-RSH-CAM is considered here.}
\end{figure}
%

\begingroup
\begin{table}
\caption{\label{tab1} Screening parameters, $\mu$ (in bohr $^{-1}$) and band gap energies, $E_g$ (in eV) of the $32$ semiconductors and insulators calculated using different dielectric constants. Band gaps are corrected for spin-orbit coupling as mentioned in ref.~\cite{WeiGiaRigPas2018}. MAE of screening parameters (with respect to DD-RSH-CAM) and band gaps (with respect to Expt.+ZPR) are in bohr $^{-1}$ and eV, respectively. }
\begin{tabular}{cccccccccccccccccccccccccccccccccc}
\hline\hline
&\multicolumn{2}{c}{$\mu$ (bohr$^{-1}$)}&\multicolumn{4}{c}{E$_g$ (eV)} 
\\ 
\cline{2-3} \cline{5-7} 
	Solids&Ref.~\cite{WeiGiaRigPas2018} &	$\mu_{eff}^{fit}$	&& Cal.$^a$&Cal.$^b$	&Expt.+ZPR~\cite{WeiGiaRigPas2018}& 
	\\
	\hline
Al$_2$O$_3$	&	0.80	&	0.84	&	&	9.76	&	9.87	&	9.10	\\
AlAs	&	0.63	&	0.73	&	&	2.10	&	2.17	&	2.28	\\
AlN	&	0.77	&	0.83	&	&	6.25	&	6.23	&	6.47	\\
AlP	&	0.66	&	0.78	&	&	2.43	&	2.39	&	2.54	\\
Ar	&	0.74	&	0.51	&	&	14.90	&	15.19	&	14.30	\\
BN	&	0.89	&	1.02	&	&	6.45	&	6.49	&	6.74	\\
BP	&	0.75	&	0.94	&	&	1.95	&	1.91	&	2.10	\\
C	&	0.90	&	1.09	&	&	5.59	&	5.59	&	5.85	\\
CaO	&	0.78	&	0.75	&	&	7.21	&	7.07	&	7.09	\\
CdS	&	0.66	&	0.67	&	&	2.79	&	2.65	&	2.64	\\
CdSe	&	0.62	&	0.64	&	&	1.86	&	1.84	&	1.88	\\
Cu$_2$O	&	0.70	&	0.72	&	&	2.61	&	2.24	&	2.20	\\
GaAs	&	0.63	&	0.71	&	&	1.28	&	1.34	&	1.57	\\
GaN	&	0.75	&	0.80	&	&	3.58	&	3.51	&	3.68	\\
GaP	&	0.66	&	0.75	&	&	2.30	&	2.26	&	2.43	\\
Ge	&	0.62	&	0.71	&	&	0.58	&	0.62	&	0.79	\\
In$_2$O$_3$	&	0.60	&	0.74	&	&	3.82	&	3.65	&	2.93	\\
InP	&	0.63	&	0.70	&	&	1.41	&	1.35	&	1.47	\\
LiCl	&	0.70	&	0.70	&	&	9.69	&	9.69	&	9.57	\\
LiF	&	0.82	&	0.69	&	&	15.86	&	16.35	&	15.35	\\
MgO	&	0.80	&	0.81	&	&	8.25	&	8.31	&	8.36	\\
MoS$_2$	&	0.73	&	0.82	&	&	1.33	&	1.30	&	1.29	\\
NaCl	&	0.69	&	0.62	&	&	8.87	&	8.90	&	9.14	\\
NiO	&	0.82	&	0.82	&	&	4.91	&	4.21	&	4.30	\\
Si	&	0.65	&	0.75	&	&	1.07	&	1.04	&	1.23	\\
SiC	&	0.77	&	0.93	&	&	2.41	&	2.39	&	2.53	\\
SiO$_2$	&	0.73	&	0.67	&	&	11.00	&	11.16	&	9.70	\\
SnO$_2$	&	0.75	&	0.76	&	&	3.62	&	3.62	&	3.60	\\
TiO$_2$	&	0.76	&	0.77	&	&	3.90	&	3.57	&	3.45	\\
ZnO	&	0.75	&	0.72	&	&	4.07	&	3.66	&	3.60	\\
ZnS	&	0.69	&	0.71	&	&	4.06	&	3.94	&	3.94	\\
ZnSe	&	0.65	&	0.68	&	&	2.90	&	2.92	&	2.87	\\
\hline
MAE &&0.07$^c$&&0.29$^d$&0.25$^d$\\
\hline\hline
\end{tabular}
\begin{flushleft}
a) calculated using $\epsilon_\infty^{-1}$ obtained from DD-RSH-CAM and $\mu_{eff}^{fit}$.\\
b) calculated using $\epsilon_\infty^{-1}$ obtained from RPA@PBE and $\mu_{eff}^{fit}$.\\
c) MAE in bohr$^{-1}$ with respect to the $\mu$ calculated in Ref.~\cite{WeiGiaRigPas2018} for DD-RSH-CAM.\\
d) MAE in eV with respect to experimental.
\end{flushleft}
\end{table}

\begingroup
\begin{table}
\caption{\label{tab1a} Mean positions of the occupied $d$ band (in eV) relative to the VBM for selective semiconductors. The theoretical values are calculated by averaging the $d$ state
energies at the $\Gamma$ point.}
\begin{tabular}{cccccccccccccccccccccccccccccccccc}
\hline\hline
	Solids& &Calculated$^a$& &Calculated$^b$	&$GW\Gamma^1$@HSE06$^d$ &Expt.$^c$	\\
	\hline
CdS	&	&	9.9	&	&	10.0	& 9.5	&	9.6	\\
CdSe	&	&	10.3	&	&	10.2	&	 9.7&	10	\\
InP	&	&	17.1	&	&	17.1	&16.9	&	16.8	\\
GaAs	&	&	20.3	&	&	20.3	&	18.5&	18.9	\\
GaN	&	&	18.1	&	&	18.1	&17.0	&	17.0	\\
GaP	&	&	19.8	&	&	19.8	&18.3	&	18.7	\\
ZnO	&	&	7.3	&	&	7.9	& 7.1	&	7.5	\\
ZnS	&	&	9.5	&	&	9.8	&8.4	&	9.0	\\
ZnSe	&	&	10.1	&	&	10.2	&8.6	&	9.2	\\
\hline
MAE &&0.7&&0.7&0.3&\\
\hline\hline
\end{tabular}
\begin{flushleft}
a) calculated using $\epsilon_\infty^{-1}$ obtained from DD-RSH-CAM and $\mu_{eff}^{fit}$.\\
b) calculated using $\epsilon_\infty^{-1}$ obtained from RPA@PBE and $\mu_{eff}^{fit}$.\\
c) See Table VI of ref.~\cite{WeiGiaRigPas2018} for experimental values.\\
d) From ref.~\cite{AndrKressHinu2014}.
\end{flushleft}
\end{table}

\begin{table*}
\caption{Ionization potentials (IPs) and electron affinities (EAs) (where EA=IP-E$_g$ in eV) of II-VI and III-V semiconductors as obtained from different methods.}
\begin{ruledtabular}
\begin{tabular}{cccccccccccccccccccccccccccc}
&\multicolumn{2}{c}{PBE}&\multicolumn{2}{c}{HSE06~\cite{AndrKressHinu2014}}&\multicolumn{2}{c}{DD-RSH-CAM} ($\mu_{eff}^{fit}$)$^a$&\multicolumn{2}{c}{$GW\Gamma^1$@HSE06~\cite{AndrKressHinu2014}}&\multicolumn{2}{c}{Expt.~\cite{AndrKressHinu2014}} 
\\ 
\cline{2-3} \cline{4-5}\cline{6-7}\cline{8-9}\cline{10-11}\\
 Solids&IP&EA&IP&EA&IP&EA&IP&EA&IP&EA

  \\
\hline
AlAs	&	5.25	&	3.86	&	5.19	&	3.09	&	5.52	&	3.42	&	5.97	&	3.52	&	5.66	&	3.50	\\
AlP	&	5.71	&	4.11	&	5.65	&	3.36	&	6.03	&	3.60	&	6.40	&	3.68	&	6.43	&	3.98	\\
CdS	&	5.97	&	4.96	&	6.56	&	4.37	&	6.55	&	3.76	&	6.94	&	4.35	&	6.10	&	3.68	\\
CdSe	&	5.60	&	5.24	&	6.06	&	4.48	&	5.99	&	4.12	&	6.62	&	4.71	&	6.62	&	4.50	\\
GaAs	&	4.87	&	4.28	&	5.16	&	3.73	&	4.82	&	3.54	&	5.38	&	3.85	&	5.59	&	4.07	\\
GaP	&	5.50	&	3.97	&	5.74	&	3.43	&	5.63	&	3.33	&	5.85	&	3.32	&	5.91	&	3.65	\\
Ge	&	4.69	&	4.62	&	4.60	&	3.78	&	4.33	&	3.75	&	4.98	&	4.15	&	4.74	&	4.00	\\
InP	&	5.15	&	4.74	&	5.60	&	4.10	&	5.10	&	3.69	&	5.74	&	4.16	&	5.77	&	4.35	\\
Si	&	4.89	&	4.30	&	5.21	&	4.04	&	4.97	&	3.90	&	5.46	&	4.11	&	5.22	&	4.05	\\
ZnS	&	6.09	&	4.11	&	6.74	&	3.42	&	7.13	&	3.06	&	7.18	&	3.29	&	7.50	&	3.90	\\
ZnSe	&	5.61	&	4.61	&	6.15	&	3.71	&	6.44	&	3.54	&	6.71	&	3.79	&	6.79	&	4.09	\\
	&		&		&		&		&		&		&		&		&		&		\\
MAE (eV)	&	0.63	&	0.51	&	0.42	&	0.33	&	0.43	&	0.38	&	0.21	&	0.28	&		&		\\
MAPE	&	9.96	&	12.92	&	6.60	&	8.54	&	7.18	&	9.43	&	3.64	&	7.07	&		&		\\
\end{tabular}
\end{ruledtabular}
\begin{flushleft}
a) Calculated using $\epsilon_\infty^{-1}$ obtained from DD-RSH-CAM and $\mu_{eff}^{fit}$.\\
\end{flushleft}
\label{tab-ip}
\end{table*}

\begin{table*}[htbp]
\caption{Lattice constant (a$_0$) and bulk modulus (B$_0$) for selective solids. MAEs of lattice constants and bulk moduli are in \AA~and GPa, respectively.}
\begin{ruledtabular}
\begin{tabular}{cccccccccccccccccccccccccccc}
&\multicolumn{3}{c}{a$_0$ (\AA)}&\multicolumn{4}{c}{B$_0$ (GPa)} 
\\ 
\cline{2-4} \cline{5-7}\\
 Solids&Calculated$^a$&Calculated$^b$&Expt.&Calculated$^a$ &Calculated$^b$&Expt.
  \\
\hline
AlP  	&	5.477	&	5.478	&	5.445$^c$	&	86.8	&	86.5	&	87.4$^d$	\\
AlAs	&	5.677	&	5.678	&	5.646$^c$	&	73.1	&	72.9	&	75.0$^d$	\\
BN   	&	3.592	&	3.591	&	3.585$^c$	&	399.7	&	416.4	&	410.2$^d$	\\
BP	&	4.531	&	4.532	&	4.520$^c$	&	165.0	&	168.3	&	168.0$^d$	\\
CdS  	&	5.908	&	5.911	&	5.808$^c$	&	58.8	&	58.4	&	64.3$^d$	\\
CdSe 	&	6.163	&	6.169	&	6.042$^c$	&	49.4	&	48.9	&	55.0$^d$	\\
C    	&	3.551	&	3.551	&	3.544$^c$	&	454.8	&	454.8	&	454.7$^d$	\\
CaO	&	4.770	&	4.771	&	4.787$^c$	&	121.2	&	120.8	&	110.0$^c$	\\
GaP  	&	5.458	&	5.460	&	5.435$^c$	&	87.0	&	86.6	&	89.6$^e$	\\
GaAs 	&	5.657	&	5.660	&	5.637$^c$	&	71.5	&	71.0	&	76.7$^e$	\\
Ge	&	5.679	&	5.684	&	5.639$^c$	&	68.4	&	67.6	&	77.3$^e$	\\
InP  	&	5.928	&	5.939	&	5.856$^c$	&	70.5	&	67.1	&	72.0$^e$	\\
LiCl	&	5.068	&	5.067	&	5.072$^c$	&	35.0	&	35.0	&	38.7$^e$	\\
LiF  	&	3.920	&	3.916	&	3.960$^c$	&	86.1	&	86.7	&	76.3$^e$	\\
MgO  	&	4.148	&	4.148	&	4.186$^c$	&	182.8	&	183.1	&	169.8$^e$	\\
NaCl	&	5.554	&	5.553	&	5.565$^c$	&	27.3	&	27.3	&	27.6$^e$	\\
SiC  	&	4.352	&	4.352	&	4.340$^c$	&	229.7	&	229.7	&	229.1$^e$	\\
Si   	&	5.444	&	5.449	&	5.415$^c$	&	95.0	&	91.0	&	100.8$^e$	\\
ZnS  	&	5.455	&	5.457	&	5.399$^c$	&	72.0	&	71.5	&	77.2$^d$	\\
ZnSe 	&	5.682	&	5.685	&	5.658$^c$	&	63.4	&	63.4	&	64.7$^d$	\\
	&		&		&		&		&		&		\\
MAE	&	0.035	&	0.037	&		&	4.8	&	4.8	&		\\
MAPE	&	0.652	&	0.693	&		&	5.3	&	5.3	&		\\
\end{tabular}
\end{ruledtabular}
\begin{flushleft}
a) Calculated using $\epsilon_\infty^{-1}$ obtained from DD-RSH-CAM and $\mu_{eff}^{fit}$.\\
b) Calculated using $\epsilon_\infty^{-1}$ obtained from RPA@PBE and $\mu_{eff}^{fit}$.\\
c) Taken from Ref.~\cite{HaasTranBlaha2009}, where the experimental lattice constants were corrected for zero-point anharmonic expansion (ZPAE).\\
d) ZPAE corrected bulk moduli values from Ref.~\cite{ZhanReillTkat2018}.
e) ZPAE corrected bulk moduli values from Ref.~\cite{LaurJudiGeor2011}.
\end{flushleft}
\label{tab3}
\end{table*}

\begin{figure*}
\begin{center}
\includegraphics[width=10 cm, height = 8 cm]{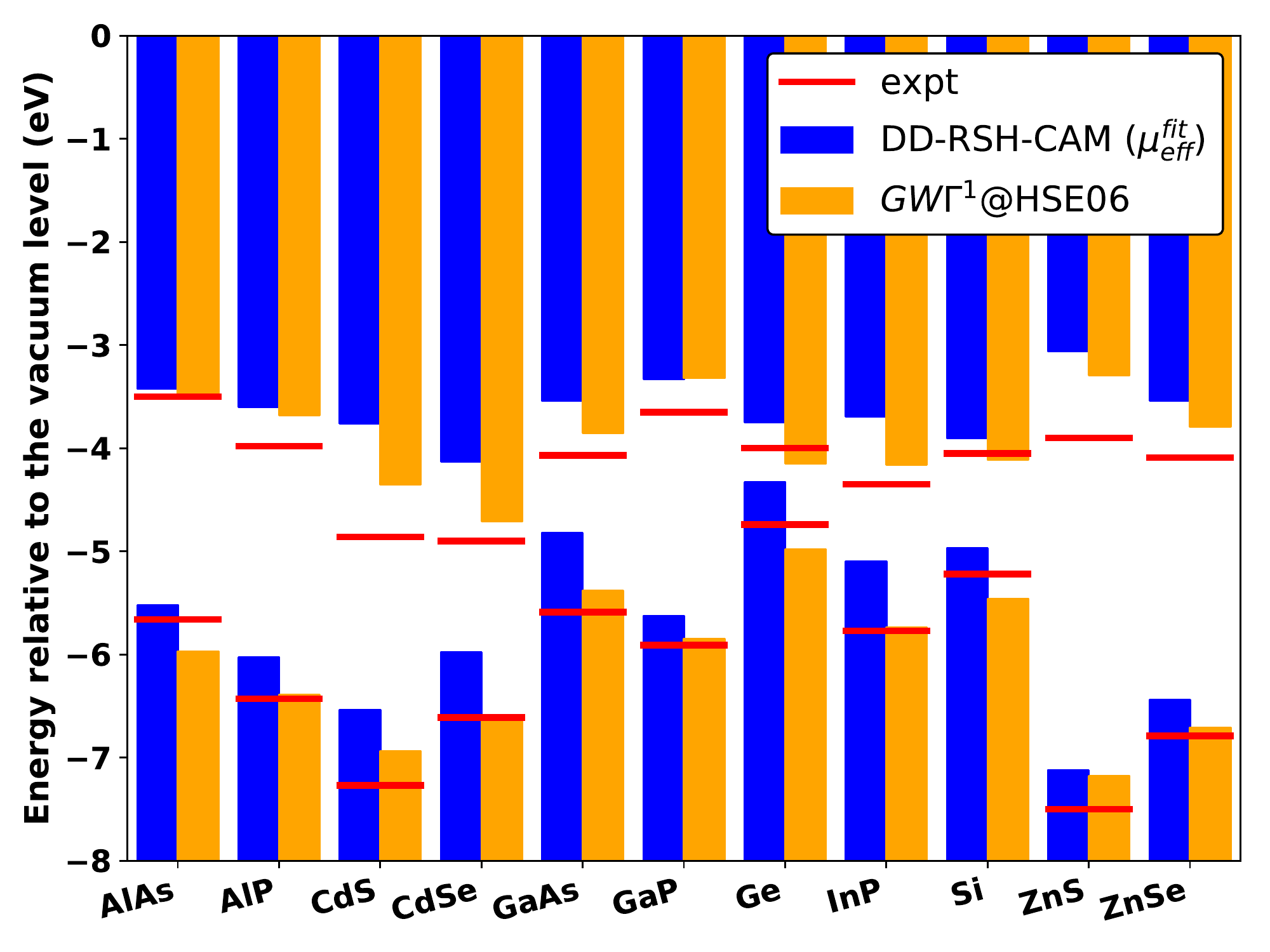}
\end{center}
\caption{\label{fig-ip-ea} Band alignment of for II-VI and III-V semiconductors based on the IPs (negatives of the VBM to the vacuum level) and EAs (negatives of CBM to the vacuum level) as obtained from DD-RSH-CAM ($\mu_{eff}^{fit}$). For comparison, $GW\Gamma^1$@HSE06~\cite{AndrKressHinu2014} (shown by the bar) and experimental values (shown by a straight line) are also shown. All the surface orientations are along (110) directions.}
\end{figure*}

We combine $\mu_{eff}^{fit}$ with DD-RSH-CAM to obtain the properties of solids. 
Unless otherwise stated DD-RSH-CAM denoted in this work is the original DDH presented in ref.~\cite{WeiGiaRigPas2018}, whereas DD-RSH-CAM($\mu_{eff}^{fit}$) corresponds to the present work.

In Fig.~\ref{fig2}, we compare $\mu$ and $\mu_{eff}^{fit}$ values for several solids, obtained from fitting with the model dielectric function and using the method described in this paper, respectively. We observe a fairly good agreement, except for a few solids like BN, C, SiC, and Ar. For the BN, C, and SiC bulk solids, $\mu_{eff}^{fit}$ overestimates over $\mu$ with less or about $\sim 0.19$ bohr$^{-1}$, while for the Ar solid, $\mu_{eff}^{fit}$ underestimates over $\mu$ with about $0.25$ bohr$^{-1}$.
One may note that larger $\mu$ values correspond to more HF mixing.

Fig.~\ref{fig3} compares the model dielectric function as a function of reciprocal lattice vector, ${\bf{G}}$. As observed from the relative mean
square error (RMSE), a maximum deviation of $0.035$ is obtained for C, whereas we observe a very good agreement for MgO. The overall analysis of Figs.~\ref{fig1} and ~\ref{fig3} suggests that the constructed $\mu_{eff}^{fit}$ from the compressibility sum rule is as good as $\mu$ obtained from the least-squares fit of the model dielectric function.

\subsection{Energy gap and valence band structure}

Table~\ref{tab1} compiles the energy gaps of $32$ semiconductors and insulators using the static dielectric constants obtained either from RPA@PBE or DD-RSH-CAM method. We consider a similar test set as the one of ref.~\cite{WeiGiaRigPas2018} for DD-RSH-CAM. One may note that the static dielectric constants that are obtained from DD-RSH-CAM are quite realistic~\cite{WeiGiaRigPas2018} (see Table TABLE IV of ref.~\cite{WeiGiaRigPas2018}). In Table~\ref{tab1} we also show a good agreement between $\mu_{eff}^{fit}$ and $\mu$ for all the considered bulk solids, and the mean absolute deviation of $\mu_{eff}^{fit}$ is $0.07$ bohr$^{-1}$ compared to $\mu$.

Next, we compare energy gaps for different solids and we observe a fairly good agreement when calculations are performed using $\varepsilon_\infty^{-1}$ of DD-RSH-CAM and $\mu_{eff}^{fit}$. One may note that in ref.~\cite{WeiGiaRigPas2018}, $\varepsilon_\infty^{-1}$ are obtained using the $f_{xc}^{bootstrap}$ kernel \cite{sharma2011bootstrap}. However, when calculations are performed with $\varepsilon_\infty^{-1}$ from the RPA@PBE method, with $\mu_{eff}^{fit}$, a slight overestimation in band gaps of insulating solids are observed (such as for Ar and LiF). This originated because for those solids RPA@PBE underestimates $\varepsilon_\infty^{-1}$, compared to DD-RSH-CAM. However, both panels of results for band gaps of Table~\ref{tab1} give an overall mean absolute error (MAE) $\sim$0.3 eV with respect to the experimental, indicating very good agreement with that of the DD-RSH-CAM~\cite{WeiGiaRigPas2018}.

Finally, Fig.~\ref{fig4} compares the experimental versus calculated band gaps for DD-RSH-CAM and DD-RSH-CAM ($\mu_{eff}^{fit}$). We observe that both methods match quite well, except few large gap solids, where slight overestimation in band gaps is observed from DD-RSH-CAM ($\mu_{eff}^{fit}$). Furthermore, band gap predictions are not very sensitive to the choice of screening parameter, $\mu$. Overall, we observe all calculated band gaps are close to that of DD-RSH-CAM~\cite{WeiGiaRigPas2018}.


Next, we calculate the mean positions of the occupied $d$ band of selective
semiconductors and the results are reported in Table~\ref{tab1a}. It is well known that approximate DFT XC functionals suffer from de-localization errors. Hence the average position of the occupied $d$ state is underestimated, even for hybrids with fixed HF percentage. As shown in ref.~\cite{WeiGiaRigPas2018}, DD-RSH-CAM can recover positions of the occupied $d$ band correctly. A very similar performance is also observed from Table~\ref{tab1a} for DD-RSH-CAM ($\mu_{eff}^{fit}$), which owns MAE of $\sim 0.7$ eV (for both cases of computing $\varepsilon_\infty^{-1}$).
These results are significantly close to that of higher-level methods such as $GW\Gamma^1$@HSE06~\cite{AndrKressHinu2014}.

\subsection{IPs and EAs}

Another serious assessment of the DDHs is to the determination of absolute band positions, hence ionization potentials (IPs) and electron affinities (EAs) calculated using the slab model~\cite{HongChen2013,JiangBlaha2016,AndrKresHinu2014,HinuAndrKresOba2014,GhoshJanaRauch2022}. As stated previously, due to the self-interaction error (SIE) (or delocalization error), semilocal functionals tend to underestimate relative band positions. Therefore, it is interesting to assess the performance of DD-RSH-CAM for extended systems, where the magnitude of SIE strongly depends on the screening nature of the material under consideration. Here, we report IPs and EAs for II-VI and III-V semiconductors using DD-RSH-CAM ($\mu_{eff}^{fit}$). We do not include DD-RSH-CAM as we expect very similar performance.

Since a direct implementation of the DDHs to the slab model is not feasible because of the high computational cost, a more trivial way of doing this is to incorporate the corrections to the VBM state of the bulk system from DDHs. Whereas, the surface supercell slab calculations are performed using semilocal LDA/GGA approximations. In the present case, we use Perdew-Burke-Ernzerhof (PBE) GGA functional. We recall, this method is similar to that of the quasi-particle (QP) $GW$-VBM approach as proposed in ref.~\cite{HongChen2013,JiangBlaha2016}. Following the protocols of $GW$-VBM method~\cite{HongChen2013,JiangBlaha2016}, the ionization potential at the DDHs level theory can also be defined as
\begin{equation}
 \rm IP^{DDH}=IP^{SL}-\Delta\varepsilon_{VBM}^{DDH}~.
\end{equation}
Here, $\rm IP^{SL}$ is the IPs calculated in the semilocal level (LDA/GGA) as follows,
\begin{eqnarray}
 \rm IP^{SL}=[\epsilon_{\mathrm{Vac,s}}-\epsilon_{\mathrm{Ref,s}}]-[\epsilon_{\mathrm{VBM,b}}-\epsilon_{\mathrm{Ref,b}}]~,
\end{eqnarray}
 and the corrections or shift to the VBM of the bulk solid because of DDH ($\Delta\varepsilon_{VBM}^{DDH}$)  is  given by,
\begin{eqnarray}
 \rm \Delta\varepsilon_{VBM}^{DDH}=[\epsilon_{\mathrm{VBM,b}}^{DDH}-\epsilon_{\mathrm{Ref,b}}^{DDH}]-[\epsilon_{\mathrm{VBM,b}}-\epsilon_{\mathrm{Ref,b}}]~.
\end{eqnarray}
Here, $\epsilon_{\mathrm{Vac,s}}-\epsilon_{\mathrm{Ref,s}}$ is calculated for the surface supercell from semilocal functionals, which is PBE for the present case. For both the zincblende ($zb$) and diamond structures, we construct the surface supercell along the (110)
direction. $\rm \epsilon_{\mathrm{Vac,s}}$ and
$\rm \epsilon_{\mathrm{Ref,s}}$ are the
macroscopic average of the local electrostatic potential in the vacuum
and the bulk region of the supercell, respectively. From bulk calculations of semilocal and DDHs, $\rm \epsilon_{\mathrm{VBM,b}}-\epsilon_{\mathrm{Ref,b}}$ and $\epsilon_{\mathrm{VBM,b}}^{DDH}-\epsilon_{\mathrm{Ref,b}}^{DDH}$ are determined, with
$\epsilon_{\mathrm{VBM,b}}$ (or $\epsilon_{\mathrm{VBM,b}}^{DDH}$) being the position of VBM in semilocal (or DDH) and $\epsilon_{\mathrm{Ref,b}}$ ($\epsilon_{\mathrm{Ref,b}}^{DDH}$) is the reference level
for the bulk calculation for semilocal (or DDH), i.e., the average of the electrostatic potential in the unit cell. We show IPs and EAs of
DD-RSH-CAM ($\mu_{eff}^{fit}$) along with the HSE06 and $GW\Gamma^1$@HSE06 in Table~\ref{tab-ip}. The VBM position (calculated from IPs and EAs) from DD-RSH-CAM ($\mu_{eff}^{fit}$) are also shown in Fig.~\ref{fig-ip-ea} along with experimental IPs and EAs. As shown in Table~\ref{tab-ip} and Fig.~\ref{fig-ip-ea}, we observe that for II-VI and III-V semiconductors, IPs and EAs as obtained from DD-RSH-CAM ($\mu_{eff}^{fit}$) are well respected and have similar accuracy with the HSE06 ones. One may note that for II-VI and III-V solids, the performance of HSE06 is respectable~\cite{AndrKresHinu2014}, as those are medium-range band gap solids and HSE06 describes well their screening. The similar accuracy of DD-RSH-CAM ($\mu_{eff}^{fit}$) indicates that DD-RSH-CAM ($\mu_{eff}^{fit}$) functional might also be a good choice for electronic structure calculations of semiconductors defects~\cite{MichPeteThom2020,SteiEiseHele2010,NguyColoFerr2018,DeakLorkeAradi2019}, where HSE06 with fixed mixing parameter is not sufficient~\cite{DeakLorkeAradi2019}.
\begin{figure}
\begin{center}
\includegraphics[width=\columnwidth]{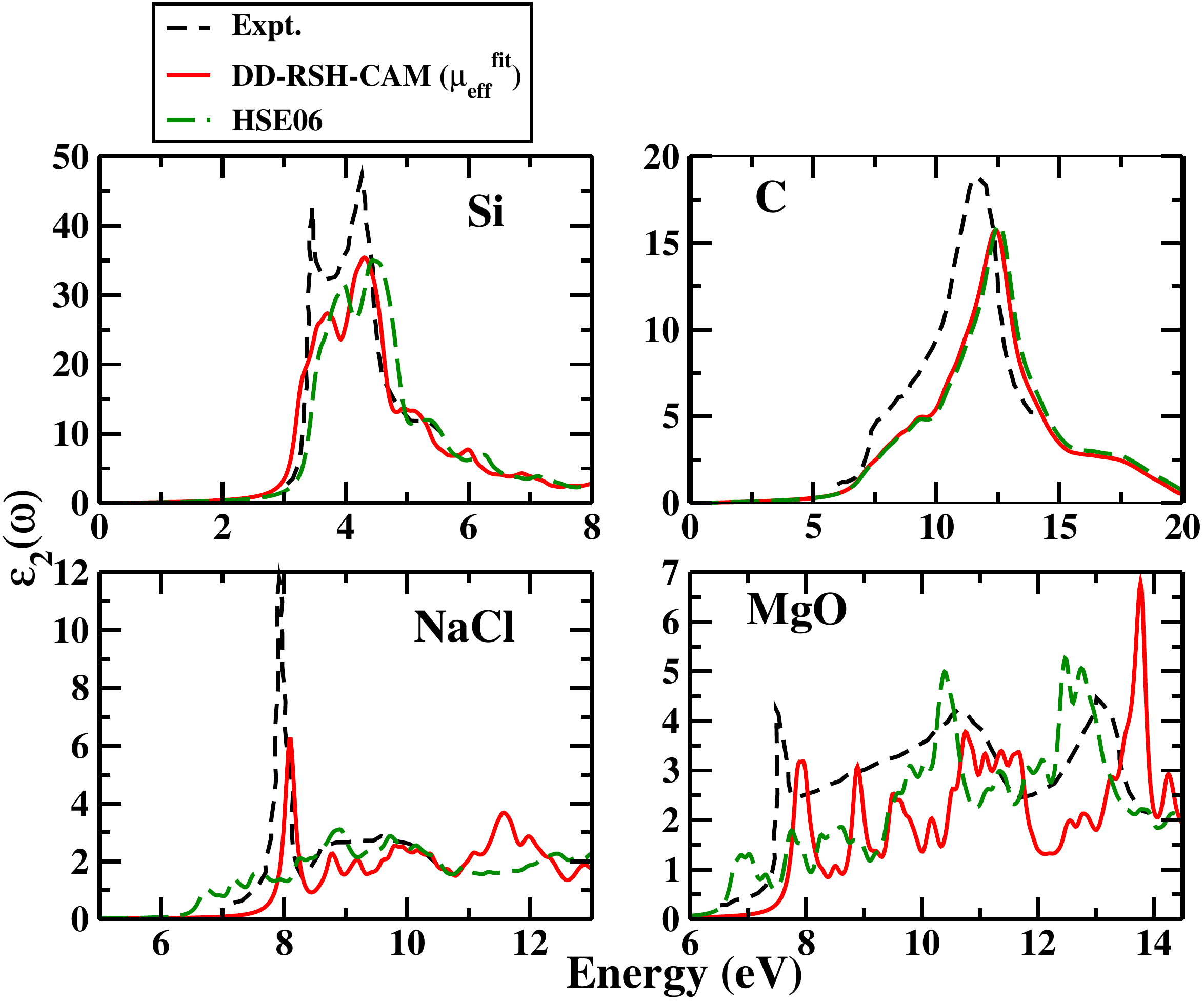}
\end{center}
\caption{\label{fig-opt} The absorption spectra of Si, C, NaCl, and
MgO, calculated with DD-RSH-CAM ($\mu_{eff}^{fit}$) and HSE06. Experimental spectra are taken from Ref.~\cite{BIRKEN1998279} (for Si), Ref.~\cite{LogoLautCard1986} (for C), Ref.~\cite{RoesWalk1968} (for NaCl), Ref.~\cite{BortFrenJone1990} (for MgO). Calculations of DD-RSH-CAM ($\mu_{eff}^{fit}$) are performed with DD-RSH-CAM's $\epsilon_\infty^{-1}$.}
\end{figure}
\begin{figure}
\begin{center}
\includegraphics[width=\columnwidth]{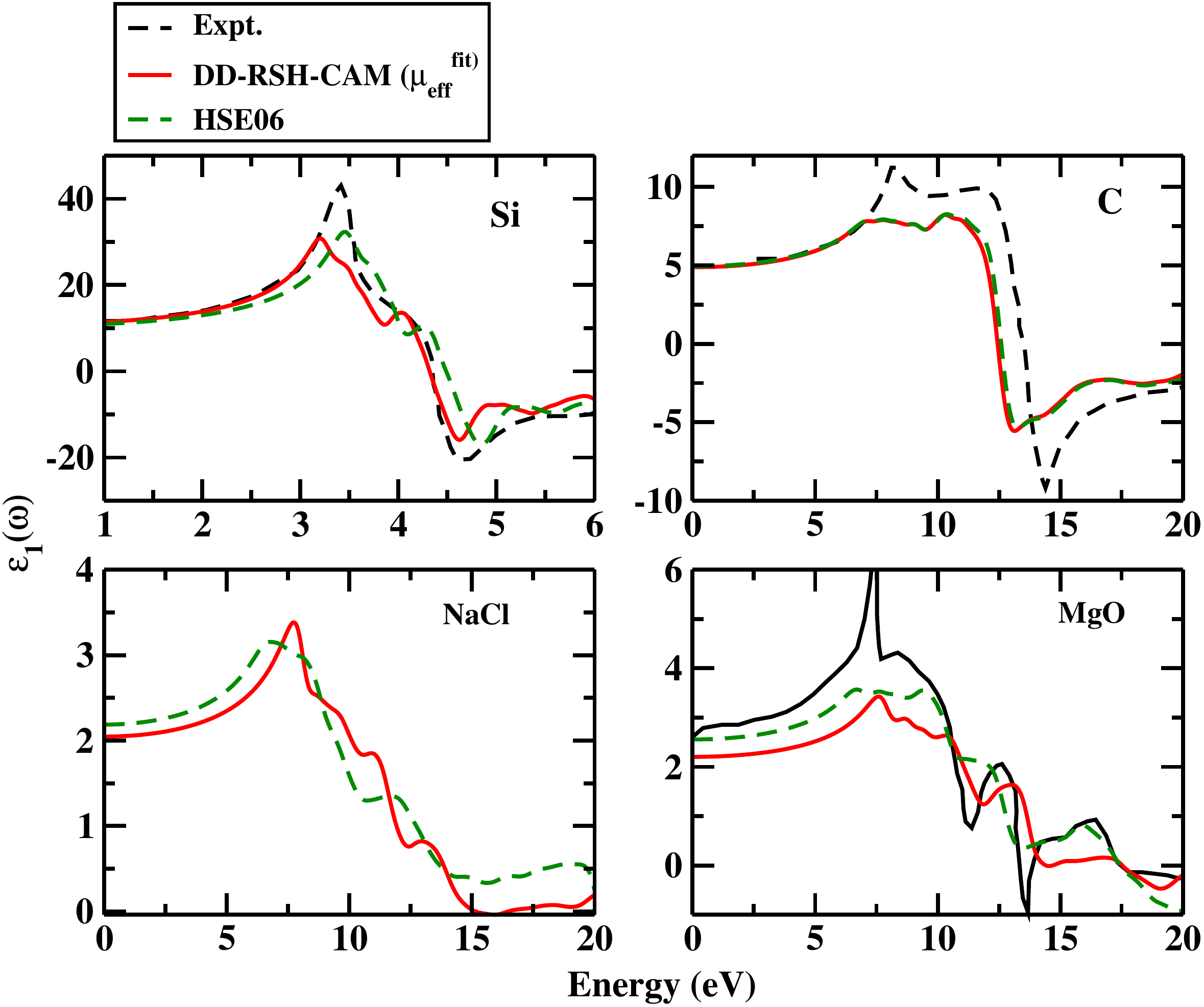}
\end{center}
\caption{\label{fig-opt-2} $\epsilon_1(\omega)$ for  Si, C, and MgO  calculated with DD-RSH-CAM ($\mu_{eff}^{fit}$) and HSE06. The Si and C experimental data are taken from Ref.~\cite{Kootstra2000application}, and the MgO experimental data is from ref.~\cite{Roessler1967electronic}. Calculations of DD-RSH-CAM ($\mu_{eff}^{fit}$) are performed with DD-RSH-CAM's $\epsilon_\infty^{-1}$.}
\end{figure}
\subsection{Optical absorption spectra}

Hybrids functionals include non-local potential, which is the key for improving optical properties of bulk solids~\cite{Ullrichtddft,paier2008dielectric,Yang2015simple,wing2019comparing,stadele1999exact,
petersilka1996excitation,kim2002excitonic,Sun2020optical,sun2020lowcost,Kootstra2000application}.
The optical absorption spectra as obtained from
DDHs are realistic~\cite{tal2020accurate}, including excitonic effects i.e., $\sim\frac{1}{\epsilon_\infty^{-1} q^2}$ in the long-wavelength limit ($q\to 0$)\cite{paier2008dielectric,wing2019comparing}. The performance of DD-RSH-CAM is studied in ref.~\cite{tal2020accurate}. Therefore, those spectra are not shown in this work.
We also recall that several low-cost XC kernels are available to compute optical properties of semiconductors and insulators~\cite{Trevisanutto2013optical,terentjev2018gradient,sharma2011bootstrap,rigamonti2015estimating,van2002ultranonlocality,
cavo2020accurate,Byun2017assessment,Byun2020time}, describing well the excitons and excitonic effects (e.g. Bootstrap~\cite{sharma2011bootstrap} and  JGM ~\cite{Trevisanutto2013optical,terentjev2018gradient} kernels), in contrast to the RPA and adiabatic LDA (ALDA) kernels.

All calculations of  DD-RSH-CAM ($\mu_{eff}^{fit}$) are performed by solving the Casida equation as mentioned in ref.~\cite{tal2020accurate}.
To assess the performance of the DD-RSH-CAM ($\mu_{eff}^{fit}$), we calculate the imaginary ($\epsilon_2(\omega)$) and real ($\epsilon_1(\omega)$) parts of the macroscopic dielectric function $\epsilon^M$ of Si, C, MgO, and NaCl in the optical limit of small wave vectors,
\begin{eqnarray}
 \epsilon_2(\omega)&=& \Im \{\lim_{q\rightarrow 0}\epsilon^M(q,\omega)\}\nonumber \\
 \epsilon_1(\omega)&=& \Re \{\lim_{q\rightarrow 0}\epsilon^M(q,\omega)\}.
\label{eq8bb}
\end{eqnarray}
We recall that the optical absorption spectrum is given by $\epsilon_2(\omega)$, while other optical properties (e.g. Fresnel reflectivity at normal incidence, and the long-wavelength
limit of the electron-energy-loss function) depend on both $\epsilon_1(\omega)$ and $\epsilon_2(\omega)$.

The optical spectra are shown in Fig.~\ref{fig-opt}. For the Si bulk, we observe quite realistic absorption spectra from DD-RSH-CAM ($\mu_{eff}^{fit}$), showing two excitation peaks at the right positions corresponding to the experimental. However, the first peak at $\sim 3.5$ eV, which represents the oscillator strength is always underestimated, similar to the hybrids with fixed mixing parameters. A similar performance is also observed for DD-RSH-CAM as shown in  ref.~\cite{tal2020accurate}.
Next, considering the optical spectra of the medium gap semiconductor C diamond, the DD-RSH-CAM ($\mu_{eff}^{fit}$) peak is blueshifted with about $1$ eV, because of the slightly larger value of $\mu_{eff}^{fit}$ (see Fig. 2). However, we obtain reasonable absorption spectra of NaCl and MgO insulators, that is considered difficult tests for all the computational methods. Compared to the absorption spectra of DD-RSH-CAM, as studied in ref.~\cite{tal2020accurate}, we see similar tendencies from the present method. One may also note from Fig.~\ref{fig-opt}, that for the wide band gap insulators, the performance of HSE06 is unsatisfactory, underestimating the absorption peak.

Furthermore, in Fig.~\ref{fig-opt-2}, we show the real part of the
dielectric function. In the cases of Si and C, both the TDDFT spectra of  DD-RSH-CAM ($\mu_{eff}^{fit}$) and HSE06 are in excellent agreement with the experimental data. However, for the NaCl and MgO insulators, DD-RSH-CAM ($\mu_{eff}^{fit}$) is more realistic, and the peaks are in the correct positions.




%
\begin{figure}
\begin{center}
\includegraphics[width=\columnwidth]{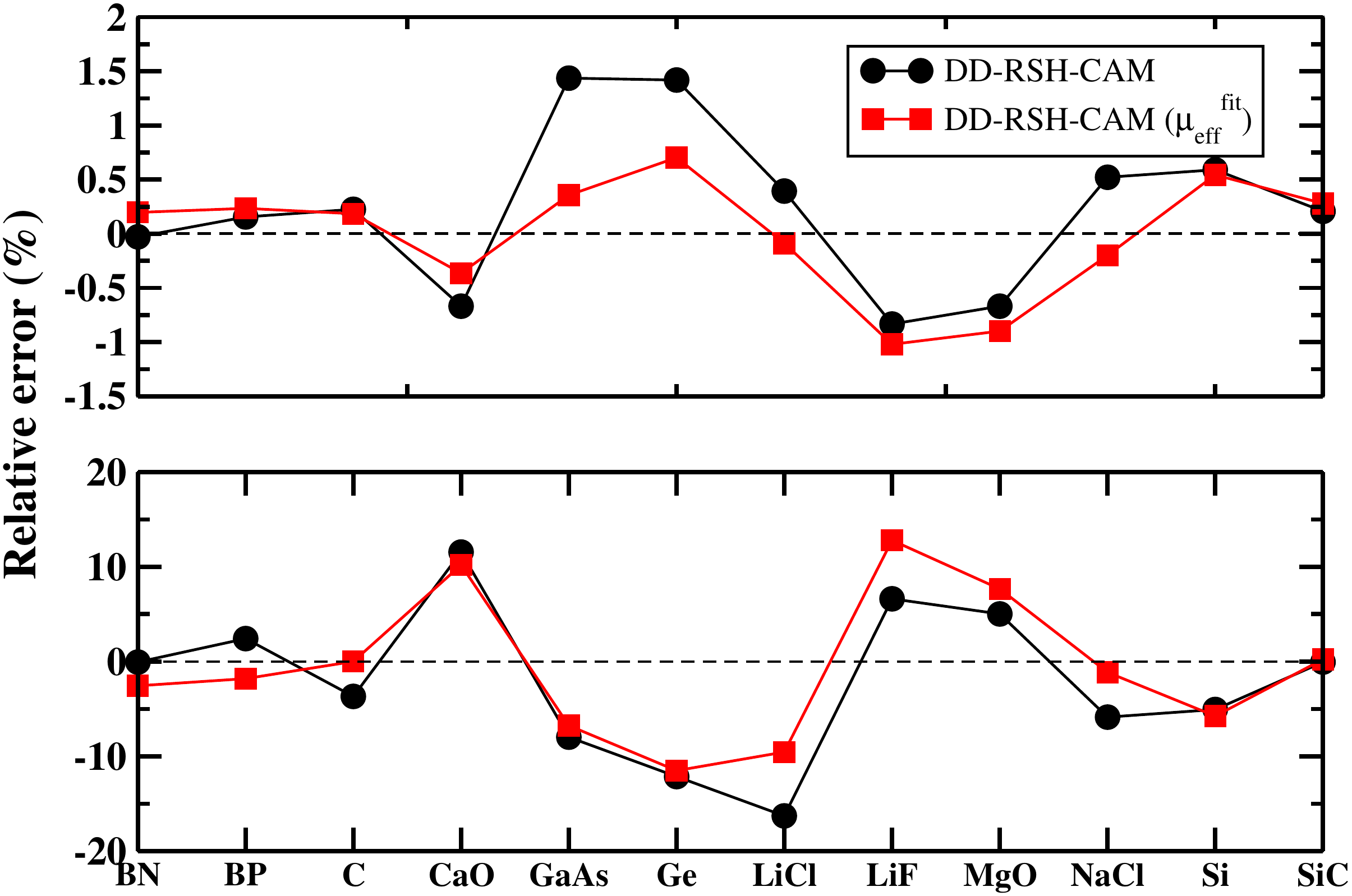}
\end{center}
\caption{\label{fig2b} DD-RSH-CAM and  DD-RSH-CAM ($\mu_{eff}^{fit}$) relative errors (in \%) for the equilibrium lattice constants (upper panel) and bulk moduli (lower panel) of several bulk solids. The DD-RSH-CAM results are from ref.~\cite{WeiGiaRigPas2018}. The calculated mean absolute percentage errors (MAPEs) of DD-RSH-CAM and  DD-RSH-CAM ($\mu_{eff}^{fit}$) for lattice constants are obtained to be 0.6\% and 0.4\%, respectively. For bulk moduli, MAPEs are 6\% for both cases.}
\end{figure}

\subsection{Structural properties}

Next, we have calculated the structural properties of selective solids with DD-RSH-CAM ($\mu_{eff}^{fit}$). Table~\ref{tab3} suggests that one can obtain quite a good performance for a wide range of solids with DD-RSH-CAM ($\mu_{eff}^{fit}$). For lattice constants, the overall MAE of $20$ solids is obtained to be $0.04$ \AA~. For comparison, we also consider DD-RSH-CAM results from ref.~\cite{WeiGiaRigPas2018}, where lattice constants and bulk moduli of $12$ solids are calculated. In Fig.~\ref{fig2b}, we plot Relative Errors (in \%) of lattice constants and bulk moduli for $12$ solids using the two DDHs. We observe that DD-RSH-CAM ($\mu_{eff}^{fit}$) improves over DD-RSH-CAM for the lattice constants of most solids. For these $12$ solids the overall MAPEs are obtained to be 0.6\% and 0.4\% from DD-RSH-CAM and DD-RSH-CAM ($\mu_{eff}^{fit}$), respectively.  For bulk moduli, the overall MAPE is within 6\% for both DDHs. Ref.~\cite{WeiGiaRigPas2018} also suggests that HSE06 offers similar accuracy as DD-RSH-CAM for both lattice constants and bulk moduli. Hence, a good description of the structural properties can be acquired from DD-RSH-CAM ($\mu_{eff}^{fit}$) across a wide variety of materials.

\section{Conclusions}
\label{seciv}

We have presented a simple and effective way of determining the screening parameter for dielectric-dependent hybrids from the compressibility sum rule combined with the linear response time-dependent density functional theory.  When applied to the bulk solids, the resultant effective screening parameter, named $\mu_{eff}^{fit}$, performs with similar accuracy for bulk solids as that obtained from the fitting with (model) dielectric function (obtained from highly accurate RPA or GW calculations) or valance electron density. Importantly, the present effective screening parameter depends only on the local Seitz radius, which is averaged over the unit cell volume of the solid, having no fitted empirical parameters. In particular, the main advantage of the present procedure is that it does not depend on the dielectric function and it can be obtained entirely from first-principle calculations for any bulk system.

Finally, our calculations show that DD-RSH-CAM ($\mu_{eff}^{fit}$) shows similar accuracy as DD-RSH-CAM for energy gaps, positions of the occupied $d$ bands, and ionization potentials. Also, DD-RSH-CAM ($\mu_{eff}^{fit}$) is quite successful for semiconductor and insulator optical properties. Simultaneously, DD-RSH-CAM ($\mu_{eff}^{fit}$) is also quite accurate for the structural properties of solids, better than its preceding variety of DD-RSH-CAM. Importantly, one can obtain the value of $\mu_{eff}^{fit}$ quite easily for any bulk solids, which in turn reduces the computational difficulty. For example, one can obtain the macroscopic static dielectric constants from PBE (using density functional perturbation theory~\cite{Baroni2001phonons,Gonze1997dynamical}) and combine them with $\mu_{eff}^{fit}$ to perform calculations for materials.

\section*{Computational details}

All calculations of dielectric-dependent hybrids are performed using the plane-wave code Vienna Ab-initio Simulation Package
(VASP)~\cite{vasp1,vasp2,vasp3,vasp4},  version 6.4.0. For all bulk band gaps and structural properties, we use $12\times 12\times 12$ Monkhorst-Pack (MP) like $\Gamma-$centered ${\bf{k}}$ points. For Al$_2$O$_3$ and In$_2$O$_3$ we reduces the ${\bf{k}}$ points to $8\times 8\times 8$. An energy cutoff of $550$ eV is used for all our calculations. In all calculations, we used PBE pseudopotentials supplied with the VASP code. Especially, for Ga, Ge, and In relatively deep Ga $3d$,
Ge $3d$, and In $4d$ pseudopotentials are used to treat valence orbitals. All band gap calculations are performed with experimental lattice constants.

For IPs, the surface calculations with PBE GGA semilocal functional is performed (with experimental lattice constants) on slabs consisting of
$14$ atomic layers ($18-39$~\AA) followed by $14$ additional vacuum layers. An energy cutoff of $550$ eV and 15$\times$15$\times$1 MP-like ${\bf{k}}$-points are used for surface calculations. The electrostatic potential used in this work is collected from the LOCPOT output file which includes the sum of the ionic potential and the Hartree potential, not
the exchange-correlation potential~\cite{vaspLOCPOTVaspwiki}. Spin-orbit coupling is also included.

For the optical absorption spectrum, we use
$32 \times 32 \times 32$ MP-like ${\bf{k}}$-points with $20$ empty orbitals. Being very
expensive, the DDHs calculations are performed in many
shifted $8 \times 8 \times 8$ grids. All calculations are performed with experimental lattice constants. We use complex shift (CSHIFT) 0.3 to smoothen the real part of the dielectric function in all our calculations.






\twocolumngrid
\bibliography{reference.bib}
\bibliographystyle{apsrev4-1}

\end{document}